# Weighted Radial Variation for Node Feature Classification

C. Andris

Massachusetts Institute of Technology

Department of Urban Studies and Planning

77 Massachusetts Avenue 10-485

Cambridge MA 02139 USA

clio@mit.edu

## Abstract

Connections created from a node-edge matrix have been traditionally difficult to visualize and analyze because of the number of flows to be rendered in a small feature or cartographic space. Because analyzing connectivity patterns are useful for understanding the complex dynamics of human and information flow that connect non-adjacent space, techniques that allow for visual data mining or static representations of system dynamics are a growing field of research. Here, we use a Weighted Radial Variation (WRV) technique to classify a set of nodes based on the configuration of their radially-emanating vector flows. Each entity's vector is syncopated in terms of cardinality, direction, length, and flow magnitude. Of course, many features, or weights, can be attributed to an edge. The WRV process unravels each star-like entity's individual flow vectors on a 0-360° spectrum, to form a unique signal whose distribution depends on the flow presence at each step around the entity, and is further characterized by flow distance and magnitude. The signals are processed with a supervised classification method that clusters entities with similar signatures or trajectories in order to learn about types and geographic distribution of flow dynamics. We use a case study of U.S. county-to-county human incoming and outgoing migration data to test our method.

Introduction

We present a novel way to classify fixed nodes in a graph or network system by their connectivity characteristics. Many network nodes are classified by their number of neighbors or degreeness, $n^{th}$ neighbor degreeness, or weighted degreeness. For example, some well-known classification schemes that build from degreeness and weighted degreeness focus on delineation spatial methods like Modularity (Ratti et al 2010), Centrality (Xu and Harriss 2008) and visual sectioning (Radil et al 2010). These methods draw from the field of physics, for example, the work of Newman (2006) on *community detection*, Newman et al (2001) on *clustering coefficients*, and Kleinberg (1999) on *hubs and authorities*.

This process An a-spatial network (like an online game network) is suitable for this method if configured in a force-directed layout like that of Fruchterman and Reingold (1991), or other snapshot-friendly graph system structure. (see Tatemura 1997 for an example). We then present a method that classifies a nodes' character based on its geometric radial edge sequence and respective values of edge lengths, and n features associated with each edge in the sequence. We then use an unsupervised classification method to typecast the unique signal. We call this method Weighted Radial Variation, or WRV. Using these classifications, we can better visualize and thus better understand the nature and dynamics of large complex flow systems.

Our case study in this paper is geographic migration, and thus our goal is to classify geographic entities in an origin/destination flow system while preserving the individual characteristics (flow magnitude, weight and direction). Next, we discuss the state of the art of geographic flow visualization and describe the *Haystack Problem*.

Much research on flow visualization is driven by the increasing availability of large flow datasets: Data from cell phone traces, traffic sensors, flight schedules and telephone records, and government digital collections are now becoming more common sources for analysis and for fields such as transportation, logistics and operations, geography and civil engineering.

Computational methods for matrix datasets have already helped researchers in these fields learn more about human and communication transactions across the built environment. The dynamics of large, multi-scale flow systems are often measured with summary factors—like a node's degree (number of neighbors), or a hub's centrality in a whole flow system (O'Kelly 1998, Xu and Harriss 2008, Jaing and Claramunt 2004).

While these methods continue to inform spatial system dynamics and benefit from cutting-edge complex network analysis (CNA) techniques, their progress has been largely unaccompanied by spatial visualization techniques. The need for good characteristic-reducing techniques for multi-featured spatial systems, like complex flows, are an important component of the static and dynamic, user-interactive, geovisualization tools that currently support visual spatial data mining. (Anselin 2005, MacEachren and Jan Kraak 1997) A recent focus towards visualizing flows (Marble 1997 and Guo 2009) and object movement (N. Andrienko et al 2000, G. Andrenko et al 2007) and space-time dynamics (Dykes and Mountain 2003) has demonstrated the benefits of these kinds of methods.

Though rendering spatial systems is a growing topic of interest, the nature of complex geographic networks' geographic flow intersections and overlaps, present a natural visualization problem, as point-to-point datasets can have many links, yielding a 'haystack' of links—from which little analysis can be performed. (Figure 1a) While more data often leads to more statistically significant results, a *Haystack Problem*, occurs when overlapping data space does not allow for simultaneous visualization and pattern recognition. In figure 1a, rays emanate prominently from U.S. states *Washington* and *Wisconsin*, only because the rays are drawn alphabetically by the software. From this, we can imagine the layers upon layers of edges underneath. An online dynamic map of U.S. domestic migration is the best known tool for analyzing spatial migration flows. (Figure 1b) This tool, however, does not allow for simultaneous selection of more than one entity, and even if this provision was enabled, the map runs the risk of becoming a dense representation with overlapping and indistinguishable entities.

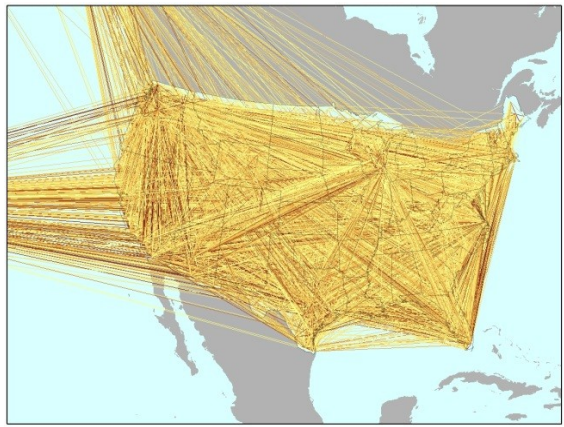

Figure 1a: Geographic flow lines often resemble a 'haystack' view of overlapping entities.

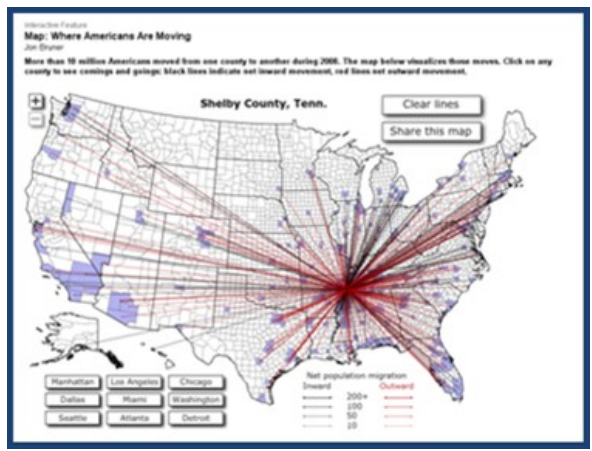

Figure 1b: An interactive flow map of migrants published by *Forbes* (Bruner, 2010) allows users to visually access migration data from a 3141 x 3141 county matrix.

Early efforts to spatialize flow dynamics include Tobler's computer mapping, where flows were rendered as aggregate arrows in order to fit in a cartographic space, and also woven into vector surfaces that resembled magnetic fields. (Tobler 1959, 1978, 1987) More recently, Andrienko and Andrienko (2010) echo the benefits of aggregating flows, adding that multi-scale analysis is now possible. Similarly, Woods et al (2010) take a unique approach to flow aggregation by assigning characteristics of OD vectors to cells, instead of the more traditional line summaries. One method that fixes the haystack problem and the aggregation problem (sidesteps summarizing or averaging values, distances or flow direction.) is an interactive system where nodes are selected, so that certain links are shown instead of all links. This provides a clearer picture of small-scale behavior, but at the expense of losing the 'all data in one view' advantage, so pre-selected views must be stored and retrieved by memory.

WRV was developed to allow analysts to better characterize nodes in a network system.

In the remainder of this paper, we introduce an example problem: understanding migration between U.S. counties. We use this problem to explain the method and illustrate how it may be used to extract novel characteristics about the network. We first introduce the data set and note some of its salient characteristics. Next, we explain the WRV method in detail. We then give a detailed characterization of the clusters of types of migration patterns identified by WRV. We then conclude with possible future applications of WRV.

## Dataset

We use data on U.S. county-to-county migration, collected by the IRS, to simulate a node/edge flow system in geographic space. We start with a matrix of 3141 x 3141 entries, where each column and row represents a U.S. county centroid. Our dataset then is whittled to 3061 nodes, as some of the 3141 counties do not post any significant migration. Unlike some network datasets, we do not count self-nodes, when movers choose a new home within the same county, but note that these values are typically high for each place. We add a distance matrix, listing the distance in kilometers between each county centroid and a matrix of angles between each node pair, where the vector head is at the origin and tail at the migrant's destination. (These can, of course, be reversed, if the destination is the node in question, but here, we concentrate on one type of flows only.)

Migrants are recorded for county-to-county flows over 10 people. The distribution of migrants per node shows an S-curve, where most counties have between 100 and 1000 migrants, relatively few have 10 to 100, while a few counties have over 10,000 or 100,000 migrants.(Figure 2) This indicates regularity among counties for over 70% of the dataset. In total, the migrant dataset contains around 5,500,000 migrants. The total migrants per county node are the summation of migrants to a county from different places of origin. The degree distribution, or the number of origins sending migrants to a destination-node, shows a similar s configuration as the distribution of migrants per node. Each node has at least one flow radiating from its center and at most 620 flows. (This maximum is found in the Phoenix, Arizona area.)Here, nearly half of the counties in the system have 10 or fewer county origins from which migrants arrive. Close to 1/3 of the number of counties in the dataset have about 10 to 100 different origins, and the remainder of the counties each has over 100. Note that the upward draw of this s curve is slower than that of the migrant distribution, indicating that there are fewer anomalies or outliers in the rank-size degree distribution and so groups of counties can be considered for their participation in classes of flow edge cardinality. These distributions are important to consider because cardinality can be a heuristic for worldliness, as it can be imagined that a geographic variety of migrants may bring more diversity to a place.

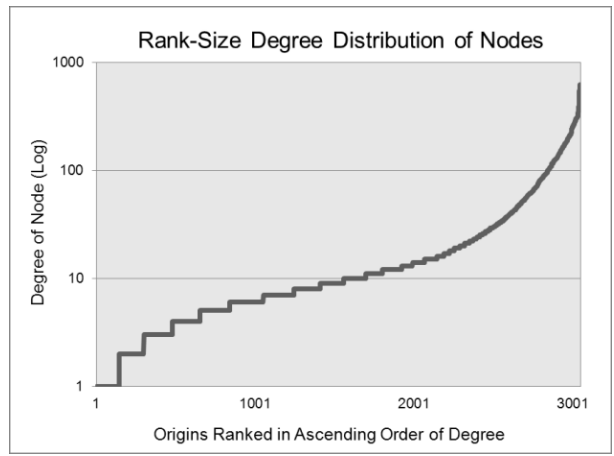

Figure 2: Rank Size Distribution of Migrants (T) and Rank Size Degree Distribution of Nodes (B) each exhibit S curves, indicating regularity towards the median values.

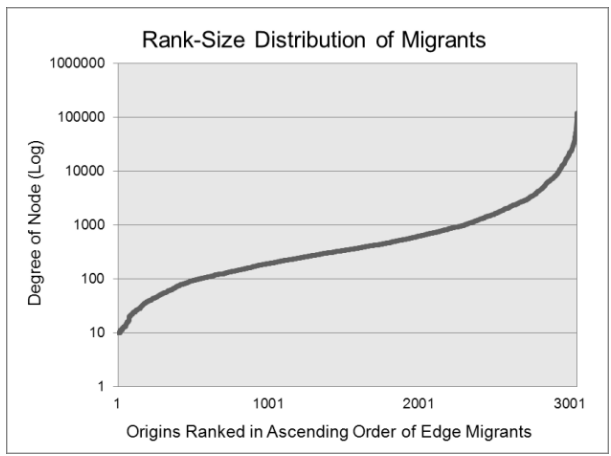

Each individual edge flow in the 'haystack' has three properties: a migrant weight (how many migrants travel on the edge), a length (the distance that the migrant travels) and an angle (the direction that the migrant travels). We explore the distribution of these statistics from an disaggregate perspective below.

We first explore the distribution of migrants per edge and find that the edge weight of migrants decreases quickly, meaning that streamlined channels are not as prevalent as edges that carry between 10 and 100 migrants. (Figure 3) The distribution of edge lengths shows that, surprisingly, 34 km is the most frequent distance travelled to migrate (541 edges participate), with a drop off at 50 km (a distance where 334 edges participate) and a final drop at around 100 km (where 150 edges participate). (Figure 4) This distribution shows that local migration is a frequent occurrence. Finally, distribution of edge angle values shows a preference for east-west migration, where edges most frequently surround peaks of 90 and 270 degrees, due east and west, respectively. (Figure 5) Interestingly, the symmetry of this distribution shows a lack of preference for a certain target migration channel.

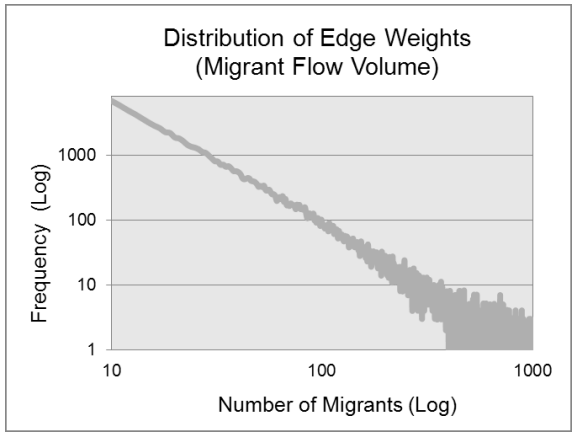

Figure 3: The distribution of edge weights for the entire system shows that a single edge rarely has more than nearly 200 migrants.

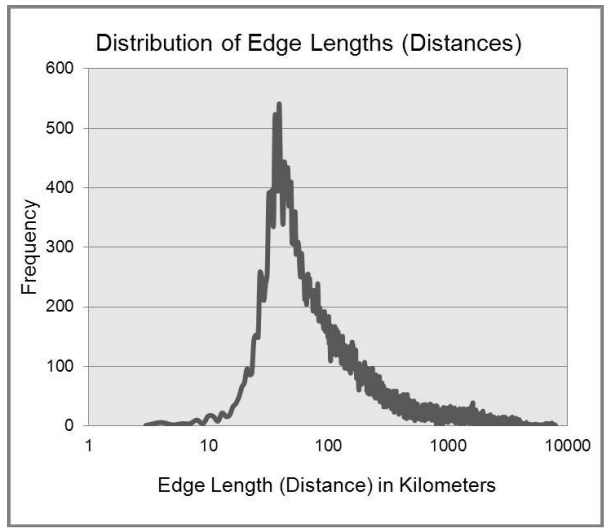

Figure 4: The distribution of edge lengths shows a generally normal configuration

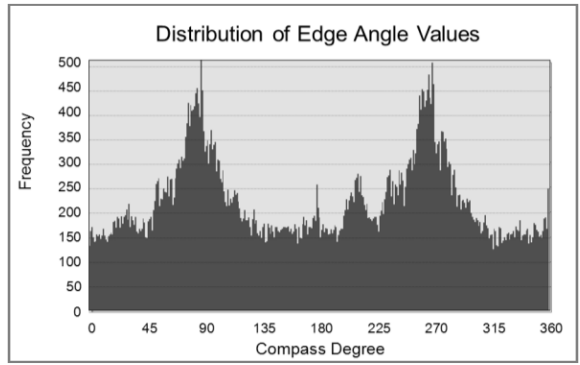

Figure 5: The distribution of edges radiating from a single node shows that the majority of moves go from east to west. Here, 0 degrees is due north.

## Methodology

Our goal is to characterize individual places by their out migration features. For each county, we extract 'stars,' where the node in focus is the center, and the flows leaving the node are attached to the central county centroid, and treated as part of the star. (Figure 6)

The star method has been used before for multidimensional data visualization, where each spoke from the center has a length equivalent to a specified quantitative feature of the entity. (Noirhomme-Fraiture 2002, Kandogan 2001) Others have taken the tool a step further, stressing interactivity (Teoh and Ma 2003) and evaluating the effectiveness of different star symbologies (Klippel et al 2009). The difference between this glyph-type entity visualization technique and our geographic case is that each vector in the radial system represents three (or more) characteristics instead of a single characteristic, as we have measures of (1) distance, (2) direction and (3) magnitude, for each. Also, our vectors are "tacked" to geographic space, meaning that an arc's radial direction is a variable with syncopated occurrences around the 0 – 360° radius, instead of an evenly-spread series of a pre-determined number of spokes in non-spatial star glyphs.

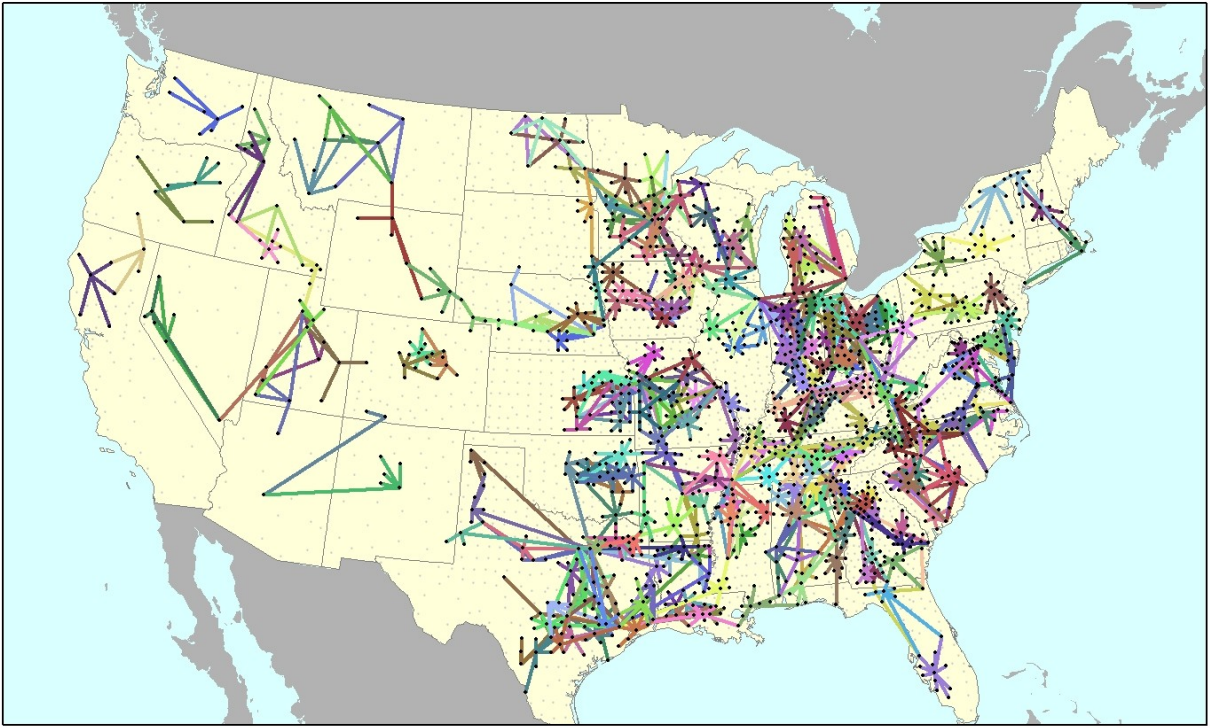

Figure 6: Selected star plots in the Continental U.S. visually represent the larger migration network system.

Noting that geographic stars (like graph structures) have a nearly infinite number of possible configurations, our probability of having a certain graph occur could be calculated by the convolution of 4 continuous variables: ray cardinality, magnitude, distance, and angle. To manage and group these e use an 'unraveling' technique to characterize the radial dynamics of each county's graph structure, to measure Weighted Radial Variation. For each county, we create a signature vector comprised of an edge weight (number of migrants) and distance value for each angle circling around the node from 0 – 360°. This signature vector is laid out as a signal over the radial steps to show similarities and differences between counties. (Figure 7) different distances (on the y axis) as the signal progresses radially from 0 to 360. These branches are also weighted by number of migrants per edge.

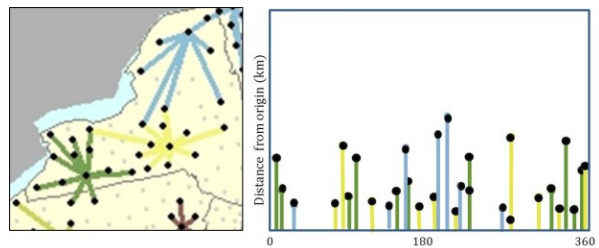

Figure 7: Three migration stars in Upstate New York (L) are 'unraveled' at right to show three individual 'signals' whose components post

*Vector Instantiation*

We create $N$ vectors, where each individual node $N_k$ is comprised of $n$ concatenated tuples $t$. In our system $N$ is equivalent to the number of counties, 3061. The number ($n$) of tuples is equivalent to the node's degree (figure 2), and ranges, as mentioned, from 1 to over 600. Additionally, each tuple $t$ is comprised of only 3 values: Differential Angle, Distance (KM), and Number of Migrants. This value sequence is chained $n$ times to form a vector, so that our shortest vector has three values (many examples), and our longest vector (representing the Phoenix area) has 620 * 3, or 1860 values

*Implementation*

Angles and distance for each edge are calculated in the ArcGIS environment to preserve geodesic distance of vector components. The weighted signals are then clustered using an unsupervised K-means clustering algorithm in the Rapid Miner Environment (Mierswa et al 2006). From these classes, a typology is created for each cluster, where the number of desired resultant cluster classes can be chosen. (Jain et al 1999) This type of signal processing and pattern recognition analysis has been successful in a geographic context for classifying and understanding space. (Reades et al 2009) We choose 10 classes for differentiating between place types.

**Results**

When these typologies of county graph types are visualized in geographic space, we are able to compare which counties are similar in their migration behavior, look at regional variation, and join demographic information. We also find that this method is robust with respect to including many edge weights per flow instead of a single measure of total migrants from county i to county j. For example, we can use our method to cluster stars where each flow has a measure of female migrants and male migrants, or migrants by age group. These typologies are then visualized via a single-variable cartographic representation that still represents the anisotropic 'spread/reach' of people migrating from different kinds of locales. From this representation, we can answer questions that would be difficult to answer otherwise, for example: how far flows travel, to what geographic direction are the flows traveling, and the magnitude of movers from each locale.

Overall, the results differentiate cities from rural areas, and show almost no clustering or autocorrelation. In fact, unless counties are of class 2, two adjacent counties are surprisingly unlikely to fall into the same class. These results lead us to believe that hierarchical drivers are in play, where k-means classes have less to do with regionalization, but with metropolitan area designation, population and population density. At first glance, we may guess that these classes are divided by such variables, as cities are certainly highlighted in

Weighted Radial Variation for Node Classification

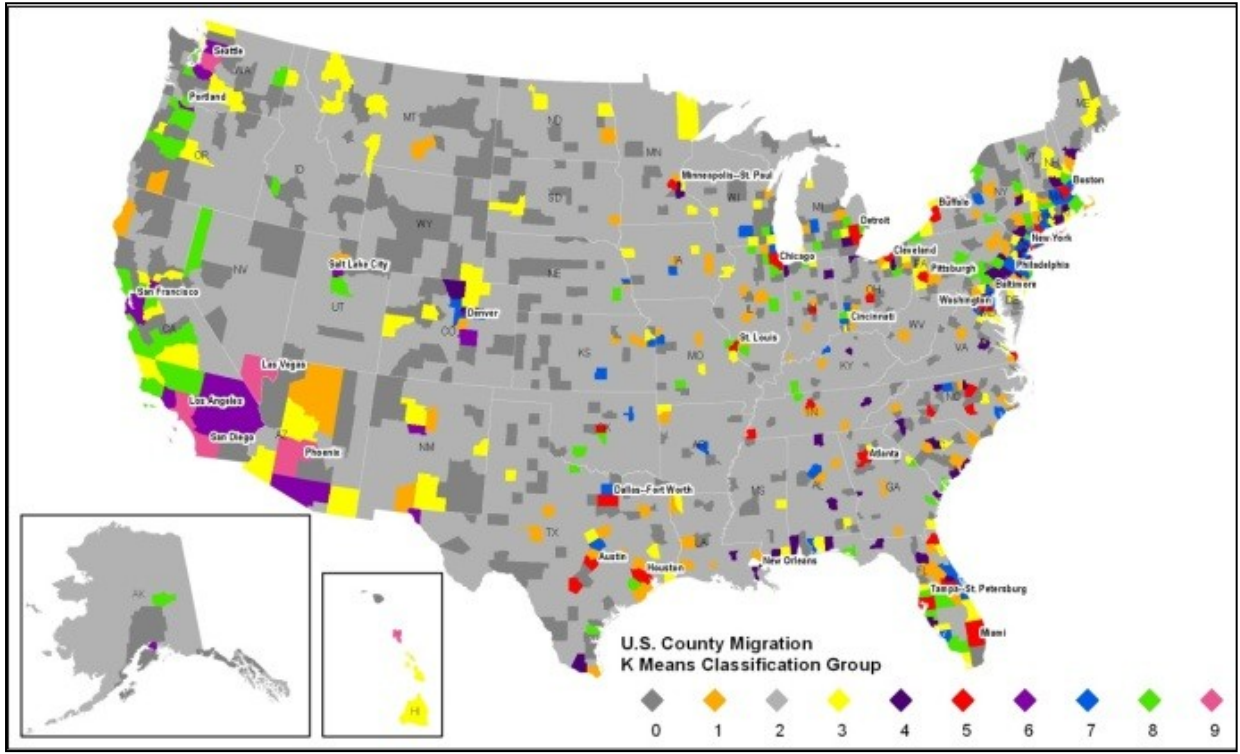

Figure 8: A map of each county by K Means class shows that geographic distribution of migration classes is dispersed, with nearly no adjacent classes.

figure 8. However, population, number of migrants and total distance traveled are not factors that can be used as singular discriminants for splitting these data (Figure 8) leading us to believe that underlying graph structure and the weighted dynamics of these graph structures are significant components of the K-means clustering assignment schedule. Below, summary statistics for each class also do not offer clear-cut discriminants, and so we focus on graph structure patterns. (Table 2) Class 1's cities draw migrants from more areas, usually in three or four main directions. This 'claw' type of structure indicates that the place is a destination for a few nearby pockets of places. In many cases counties in class one attract migrants from a number of smaller areas that are likely to be on a similar highway system. Cities in this class are similarly sized to those in class 0. Anchor cities that likely draw migrants from smaller proximal towns include Shreveport LA, Chattanooga TN, Hampton VA, Green Bay WI, Billings MT, Topeka KS. Unlike cities found in suburban outskirts and sprawl, these cities are markedly not part of urban agglomerates, nor as synergistic cities with a nearby partner, and range in population from a nearby partner and range in population from 100,000 to 200,000. One interesting finding in this group is the number of state university towns. The following cities are the primary campus of the state-wide university system: Santa Fe NM, Columbia MO, Athens GA, Lawrence KS, Iowa City IO, Medford OR, Bloomington IN, Fayetteville AR, Chapel Hill NC. Fargo ND is home to North Dakota State, College Station TX is the home of Texas A&M; Ashville NC, Lafayette LA, to large satellite campuses of the state system. Other examples in this category are Wilmington NC, Bethlehem PA, Ithaca NY house large Division 1 schools East Carolina, Cornell and Lehigh Universities.

The vast majority of counties in the U.S. fall into Class 2, although the populations of these counties are generally smaller than any other group.(Table 1) The graph structure of counties in this group are the sparest and simplest of the system (Table 2), with few spokes in a line, angle, three-tier, cross or simple star configuration. Also, entities in this class have markedly lower distances and migrants that the remainder of the classes (Figure 9). Class cities include Midwestern Capitals Frankfort KY and Jefferson City MO. Class 2 also includes large cities in the Deep South, Vicksburg MS, Tupelo MS, Decatur AL, in West Virginia: Beckley WV, Wheeling WV and Virginia: Roanoke VA, Lynchburg VA, Suffolk VA, Fredericksburg VA, Fairfax VA, Manassas VA. Although Fairfax, Manassas and Fredericksburg, are all considered part of the Washington D.C. metropolitan area, with notable commuter streams, they exhibit the same migration patterns as the plethora of small cities and sparsely populated counties. With the exception of the capital cities, the aforementioned entities have historical preservation components to their cities, and small tourist industries. Eastern VA and West Virginia are known for elderly residents; this category's counties have the oldest average age of any category. This category's cities are also generally devoid of major transportation infrastructure like airports, ethnically diversity, higher education and international draw.

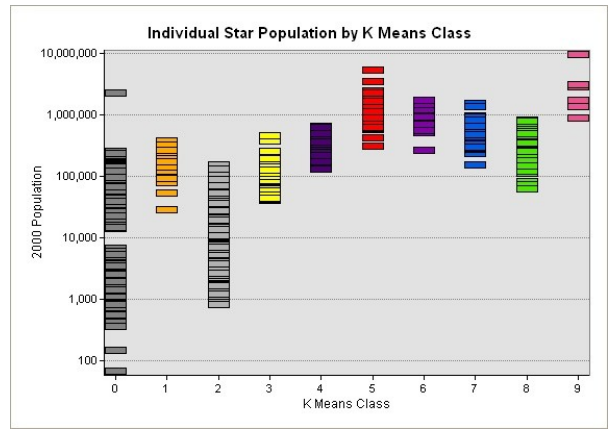

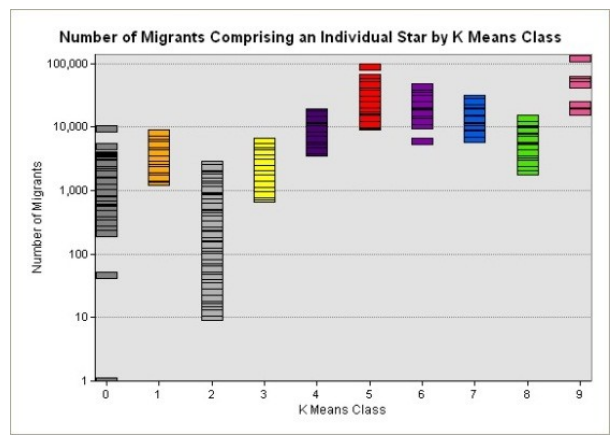

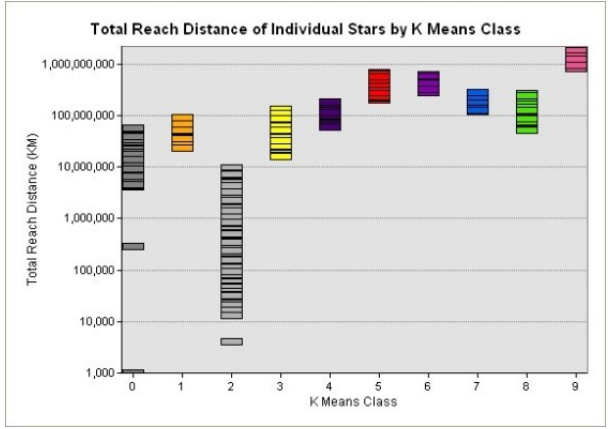

Figure 9: Plots of [Top] Total Distance, [Center] County Population and [Bottom] Migrants per Individual Stars, are on the Y axis, while each county in is plotted in its k means class. The lateral consistency in variable benchmarks for almost each class indicates that a single variable is not responsible for k means class splits.

Table 1: Entity (Star) Statistics by K Means Class

| K Means Class | Number in Class | Avg. Number of Edges | Avg. Migrants Per Star | Avg. Total Distance | Avg. County Pop (in 1000s) | Average Age |
|---|---|---|---|---|---|---|
| 0 | 394 | 21 | 1115 | 8,500 | 66 | 37.1 |
| 1 | 96 | 64 | 3384 | 46,400 | 178 | 34.6 |
| 2 | 2,297 | 8 | 320 | 900 | 24 | **37.8** |
| 3 | 105 | 43 | 2321 | 35,600 | 132 | 36.0 |
| 4 | 56 | 127 | 8031 | 120,300 | 352 | 34.9 |
| 5 | 43 | 301 | 27900 | 324,700 | 1,191 | 34.7 |
| 6 | 19 | 218 | 21065 | 408,000 | 888 | 33.8 |
| 7 | 49 | 181 | 13599 | 187,000 | 597 | 35.5 |
| 8 | 75 | 97 | 5718 | 101,300 | 303 | 34.6 |
| 9 | 7 | 465 | 51080 | 1,191,000 | 3,124 | 33.9 |
| National Average | | | | | (89) | (37.3) |

Table 2: Typologies of Common Structures by K Means Class

| Class | Common Structure | Class | Common Structure |
|---|---|---|---|
| Class 9 | | Class 4 | |
| Class 5 | | Class 1 | |
| Class 6 | | Class 3 | |
| Class 7 | | Class 0 | |
| Class 8 | | Class 2 | |

Classes 3 and 4 show distinct fanning, with class 4 exhibiting a fuller range of cities from which migrants arrive. (Table 2) Class 3 typically draws from cities in on two or three distinct directions, also indicating that road infrastructure might be in play. Cities include Lubbock TX, Kansas City KS, Greeley CO, Rochester MN, Duluth MN, Sioux City SD, Racine WI, Oshkosh WI, Kenosha WI in the Midwest, West Cheyenne WY, Missoula MT, and small cities Charlottesville VA, Dover DE, Warwick RI, Niagara Falls NY in the East. Each of the aforementioned cities is marked with a steady, stable population, and a generally homogenous citizen body. The mountain region sees a lot of class three, indicating that this type may be characteristic of sparsely populated regions. Class three also notably fills in space between coastal counties with fuller migrant draw patterns.

California cities are typically considered diverse and far-reaching, especially with respect to Asian and Hispanic populations, but Modesto CA, Davis CA, San Louis

Obispo CA fall into this category of low cardinality, and near-reaching pull. Their existence in this class may indicate that populations drawn to these cities are more standard, local and homogenous, and perhaps that the offerings of these cities are less colorful, and better-suited for attracting local community.

Cities in class 4 are marked with a wider fan of migrant origins, and are situated almost exclusively west of the Mississippi River. Class 4 counties dot the Deep South and Gulf, as well as the Northeast. Class 4 counties include larger cities in the industrial north, beginning with Baltimore MD, westward to Toledo OH, Akron OH, Lexington KY, Fort Wayne IN and St. Paul MN. St. Paul MN is often touted as the "Last City in the East", a dictum that is evidenced by its participation in this eastward-pulled class. In the Deep South and Gulf, Norfolk VA joins cities Charleston SC, Savannah GA, Huntsville AL, Montgomery AL, Gulfport, Biloxi MS, Mobile AL, New Orleans LA, in this category. Each of these cities contain older infrastructure, historical influence and less progressive policy—like relatively lax tobacco usage. As discussed by Hulten and Schwab (1984) these areas have been 'slowed by an aging public infrastructure, deteriorating urban environment an obsolete capital stock and institutional sclerosis.' (Pg. 152)

Those looking to economically invest in the temperate, cultural and aesthetically-pleasing cities like Savannah, Charleston and New Orleans may note their presence in a category with the same limited migrant magnetism as aging cities.

As an aside, this class also includes a number of large college towns: Tallahassee FL (Florida State), Gainesville FL (Florida), Baton Rouge LA (Louisiana State), Fort Collins CO (Colorado State), Trenton NJ (Rutgers), Wilmington DE (Delaware), Ann Arbor MI (Michigan), Knoxville TN (Tennessee) and Durham NC (Duke), Winston Salem NC (Wake Forest), Providence RI (Brown).

In class 5, we see the strongest pulls of any class until class 9. These cities are found exclusively in the Midwest and East, and include migrant-magnet New York NY. In concordance with each class, cities in class 5 are rarely, if ever nearby one another, but act as regional anchors throughout the Sunbelt and Snowbelt. Class 5 typically draws from the entire U.S., and therefore it is not surprising that regional "Command and Control Centers" (Frey 1990) such as Atlanta and Minneapolis-St. Paul, draw diverse populations to their "world class" influential centers. The cities distinctly dot 3 regions: the Midwestern Snowbelt region, including Buffalo and Rochester NY, Pittsburgh, Columbus OH, Cleveland, Indianapolis, St. Louis, Detroit to metropolitan Chicago; Sunbelt cities Memphis TN, Nashville TN, Charlotte NC, Raleigh NC, Tampa FL, Jacksonville FL and Texas powerhouses, Dallas, Austin and Houston, and Northeaster corridor. These designations as coherent regions are founded by previous research. For example, Rodgers (1952) defines industrial inertia in terms of The Pittsburgh-Cleveland-Buffalo triangle, marked by steel production, and its expansion to St. Louis, Indiana, Detroit and Chicago-Gary. In comparison to the cities in the Snowbelt region in class 4, these Snowbelt cities may be sidestepping economic stagnation, as they differentiate themselves with the poor-cardinality draw of class 4 cities like Fort Wayne IN. The inclusion of these classical industrial regions in the same category as warmer climates, especially the large Atlanta and Charlotte, a favorite for Fortune 500 Companies, is evidence that this 'rusty' region still has a strong similar pull, despite some discussion of its current or impending dilapidation.

Thirdly, although Boston and Washington D.C. are not explicitly in this category, their "geographical inertia" (Auty) can be seen. In the Boston area, areas rich with medical, technical, pharmaceutical, research and firms like Cambridge MA, Somerville, and Arlington, as well as D.C.'s Bethesda, and Gaithersburg MD, home of the National Institutes of Health (NIH) and National Institute of Standards and Technology (NIST) can understandably fall into this category because of the national pull for skilled workers in an area dense with higher education. Additionally, migrants to these more suburban areas are likely drawn to downtown amenities and 'worldliness', a migrant's place of residence in one of these cities can indicate that he works or reaps the benefits of the larger surrounding city.

This class is especially important because it illustrates the relative powerlessness of intervening opportunities (Stouffer 1966, Galle and Taeuber 1966, Bright and Thomas 1941) when migrants are considering these cities. Because cities in this class exhibit a strong national draw, we can insinuate that migrants cannot be satisfied with closer locales, and thus these cities have a special magnetism.

Importantly, the 'rust belt' has been cited recently as an area of economic decline. However, their participation in this group illustrates that they are still in the same class as cities that are seen as prospering.

Class 6 cities draw a variety of migrants for their relative sizes. Their social reach is diverse, with steady currents to many major cities. This class includes cities exclusively in the West: Pacific Northwest cities Tacoma WA, Everett WA, which flank Seattle, home to mega corporations Microsoft and Boeing, and Portland OR, known for its progressive urban planning and sustainability initiatives, join equally progressive Tucson AZ, and Denver CO. Most notably, California cities San Francisco, San Jose, San Bernardino, Sacramento, Oakland, Riverside CA, Fremont CA, Santa Clara, Berkeley, Oxnard, Thousand Oaks, San Mateo, Mountain View, Redlands, Redwood City, Citrus Heights, Palo Alto, Laguna, Menlo Park, and Saratoga and others fall into this group. These Silicon Valley, and satellite silicon cities are marked with high literacy and education rates, high incomes, and diverse lifestyles. This area is marked with up-and-coming research and industry, including Google, Adobe Systems, Disney Pixar, Facebook, Oracle Corporation, eBay, Apple Inc., Cisco Systems, Hewlett-Packard, Intel, Yahoo!, Sony, TiVo, Nokia, YouTube, VeriSign and Netflix. (Sturgeon 2000) This category succinctly extracts places that have the opportunity to attract and evidence of attracting from a variety of places, accordingly the average age of this class is the lowest at 33.8 (Table 1). Surprisingly, other cities in this category are Salt Lake City UT, Albuquerque NM, El Paso TX, Colorado Springs, CO and Anchorage AK, indicating that new technology firms may be sprouting here, or that site suitability models for new firms should also take these cities into account.

Class 7 cities include metropolises Philadelphia and Washington DC and a steady chain of coagulate satellites within the gravitational pull (Huff 1964) of New York City, such as Newark NJ, New Haven CT, Hartford CT, Hackensack NJ, New Brunswick NJ, White Plains NY, Levittown PA, Hicksville NY. Similarly, cities and suburbs of Boston, MA like Quincy, Lynn, Haverell, Peabody, Revere, Needham, Gloucester, Methuen, Leominster, Fitchburg, Worcester MA, are also in this class. With the exception of Needham, these proximal places, nested in the Massachusetts Bay periphery are historically known for industry, especially manufacturing.

In the Midwest, Milwaukee WI, and Cincinnati OH anchor industrial mid-western cities Grand Rapids MI, Kansas City MO, Dayton OH, and Louisville KY. Similarly-sized Birmingham AL, and Little Rock AR also join from the same longitude in the Deep South. The class continues westward to include large Great Plains cities Tulsa OK, Omaha NE, Wichita KS, and Des Moines IA.

Boulder CO, Madison WI, Annapolis MD and Syracuse NY, in this category, may attract a similar array of researchers due to Universities, research facilities and military organizations like the Naval Academy. Boulder and Annapolis are also proximal to Denver and Washington D.C., respectively, giving potential migrants the benefits of efficient air travel and access to diverse goods and services. Florida havens Pensacola, Fort Meyers and Daytona Beach are known to be warm-weather havens and may draw similar channels of aging migrants.

Class 8 migration structure is limited for the size of its populations. The structure consists of draws from two major directions, and exhibits fanning in these directions. These counties are geographically situated in the southwest and west with cities Fresno CA, Bakersfield CA, Stockton CA, and Provo UT Reno NV and Boise ID. These cities do not have the draw of larger California cities on the Pacific Coast, although interestingly, acclaimed Santa Barbara CA, considered a wealthy haven, is included in this class. Provo and Reno do not draw the same diverse migrants as their respective same state anchors Salt Lake City and Las Vegas. In a similar fashion, small Pacific Northwest cities Salem OR, Eugene OR, Vancouver WA, and Spokane WA draw a smaller, less diverse set of migrants than their local anchors, Seattle and Portland. Midwestern and Plains cities Lincoln NE, St Louis MO, Springfield MO, Lansing MI, Champaign IL are also included. Of these, St Louis is the largest city. As state capitals, Lansing and Lincoln, along with Harrisburg PA and Richmond VA, may draw intra-state citizens, like legislators and politicians. Lincoln and Champaign, along with South Bend IN, Norman OK, Binghamton NY boast large universities—three of which are comprised mostly of in-state students.

In a more temperate climate, Augusta GA, Hilton Head SC, Land O'Lakes FL, Naples FL, Holiday FL draw "snow birds" from the Midwest more than the Northeast. These cities offer amenities for the retired, especially notable golfing infrastructure.

Class 9 cities exhibit a heavy eastward draw. The opposite of class 2, this group has few counties, but each of these few counties (9) has a high cardinality of flows. Following suit, this group also draws the most migrants per city. (Table 1) The Phoenix AZ area draws the maximum flow edges at over 600. This class also includes, Las Vegas NV, Los Angeles CA and their respective large surrounding areas, as the counties in which these southwestern metropolises are situated are notably large in area. This class also includes San Diego CA, whose population is comprised of a Hispanic proportion similar to the aforementioned cities. Seattle WA and Honolulu HI, are also included here. Geographically it could have been expected that Portland and San Francisco would have been included in this class, but their structures better matched those of class 6. San Diego could have been included in class 6, but perhaps because of a lack of progressive technology, and a more traditional military hold with Naval influences, San Diego remains in the same class as its larger, sprawling cities.

Our results lead us to question what it means to have community typologies, and what can we learn from this. Even at first examination, we could tell heuristically from the cardinality of flows that some communities exhibit much extroverted outreach qualities within the flow system while others stretch less but these categories give a more colorful illustration of community typologies.

However, we stress that this technique has uses for multiple fields where node-edge graphs are use. In the field of geographical information science, the spatialization, and web-formation, of these nodes is inherent in the nature of the data. However, force-directed graphs can also benefit from this methodology.

In the next section, we discuss insights and suggest a few applications for this technique.

## Applications and Conclusion

*(1) Economic Assessments*
Using WRV, it should be possible to find out what the advantages of certain features are, or the impact of those features on the local network structure. It can be expected that a city with a world-class facility may exhibit different flow properties than its neighbors. Since the town with the special facility general holds the same characteristics as the places nearby. For example a college, or in the United States, the Mayo Clinic in Minnesota, Santa Fe Institute in New Mexico, Oak Ridge National Laboratory in Tennessee, and Cape Canaveral NASA Station in Florida are examples of facilities that draw workers and visitors from a wider variety of places, than other cities in their region.

*(2) Community Detection*
Although our results are not clustered, a clustering of entities of homogenous typologies could allow for the delineation and partitioning of similar graph communities. These classes can be a new form of community detection, if a community can be considered as nodes with similar behavior. In our case, we consider spatial similarities to delineate a geographic community or policy zone.

*(3) Prediction*

A temporal assessment of flow types could inform the changing nature of connectivity by their flow dynamics. An entity that exhibits a syncopated temporal trajectory with another entity can use the latter entity's current state as a probabilistic preview of future flow behavior, and help forsee future steady state conditions.

*(4) Behavioral Typologies*

In a group or work setting, the dynamics and form in which agents cooperate in a social system can help classify extroversion or introversion typologies for optimal partnerships. Leaders in Organization Development, Sociology and Psychology can use these models for a robust view of systemic ties.

*(5) Infrastructure Planning*

There are many potential applications for this method since this technique can take in a wide variety of flow data. Relevant data types include phone calls, emails, air transportation and commuting. Given that commuters want the most efficient way to get to work, with challenges of traffic, commuters often fight for funding for new or better infrastructure—types of infrastructure like rail, widened roads, or better airport access.

The timing of migration, and channels through which it flows, means that a model with declining moving costs fits better than a neoclassical model of free or fixed-cost mobility.

In conclusion, our aim for this new visualization technique was to preserve individual, disaggregate characteristics of flow data while allowing the information to be explored in a single view. By extracting and clustering different geographically-tacked graph configurations, we are better able to understand the distribution of human movement patterns in space, while using disaggregated data. This method is not limited to migration patterns, but can be used for other datasets where origin "reach" is a metric of interest like commuting flows, phone call volume, or temporal vacation/leisure flows.

We foresee many possible applications for WRV, both scientific and applied.

## Acknowledgement

The author acknowledges Frank Hardisty and Leon Andris for insight and assistance. This work is supported in part by the ASEE National Defense Science and Engineering Fellowship Program and the MIT Sense*able* City Lab.